# Vocalnayno: Designing a Game-Based Intervention to Support Reading Development in Primary Schools

Michael "Adrir" Scott
Brunel University, United Kingdom
michael.scott@brunel.ac.uk

**Abstract:** Encouraging children to read frequently and helping them to develop their reading skills as effectively as possible can be a challenge for some primary schools. Often, institutions have to rely on external teaching assistants and parent volunteers to provide pupils with additional one-on-one support in order help them to achieve an acceptable standard of reading. However, there are inefficiencies to this approach. For example, some volunteers lack the necessary instructional knowledge to deliver learning material effectively, while others may possess limited agency to assess pupil progress through the standard mechanisms. Consequently, this can delay the identification and communication of individual needs, which can be essential for tailoring such support.

This research questions whether the use of a game-based intervention can integrate into the existing teaching culture at a primary school and aid teaching assistants to achieve a more significant impact on pupil reading development. In order to synthesize an appropriate design, an action-research inspired approach at a local school has been adopted. This is very much a work-in-progress as several methodological challenges have already been encountered. In particular, addressing barriers to entry and including a wide variety of stakeholders in each stage of participatory design. However, the approach helps to maintain scope within relevant cultural boundaries while also addressing new weaknesses that emerge through the observation and discussion of current practices.

A prototype based on an initial process of gathering requirements is presented using Multimedia Fusion Developer 2. The design incorporates a game-like exercise where a foam volcano character releases bubbles containing letters and words. Pupils must read these aloud in order to burst them open, which is recorded as a metric of reading ability. This example could be deployed as an assessment tool on a laptop or tablet device, providing assistants with an indicator of phonetic strengths and weaknesses.

## 1. Reading in the Primary School Classroom

Reading is an important skill that is taught in primary schools in the United Kingdom (UK) as part of the Key Stage 1 (KS1) curriculum (ages 5-7). Compulsory formal assessment begins with phonics screening at age six and later involves standardised assessment tests (SATs) at age seven. Such measurements can help schools to identify whether individual pupils require extra help with the development of their reading ability. However, this is not the only means to achieve this goal. Ongoing formative assessment (Roskos & Neuman, 2012) is often encouraged within the culture of many schools because there can be large individual differences in the reading ability of new pupils. This approach enables teachers to scaffold learning material more effectively for each pupil. Often, in accordance to guidelines provided by the Local Education Authority (LEA), using a range of learning materials of specified difficulty. Unfortunately, class sizes can be large (up to 30 pupils) and often include a mixture of age groups. This means that some schools rely on parent volunteers and teaching assistants, external to the class or school, to provide one-on-one support.

While there are many benefits to involving others (Hoskisson, Sherman & Smith, 1974), there are several potential pitfalls, particularly relating to scaffolding and assessment. Helpers may not be familiar with the needs of the pupils that they are trying to help or they may not be able to provide feedback to the teacher as effectively as possible. Typically, helpers follow a learning activity with the pupil and maintain a written record of the child's progress, shown below in Figure 1. Unfortunately, this approach is open-ended and qualitative, sometimes limiting itself to a recommendation concerning the ability-band assigned to the pupil. Consequently, it may not provide potentially critical feedback on specific weaknesses, which may be of interest to the teacher. Furthermore, such information is often recorded in multiple places, such as a reading record and vocabulary book. This means that there are multiple sources of information a helper needs to manage in order to identify and record individual needs. Each, of which, can be easily missed or lost. Moreover, the system can be slow to react because the records are not always available to parents and teachers, sometimes being time-consuming to read through. The alternative of using in-class quizzes can be too narrow to accurately measure a wide-ability class and may not be particularly engaging for some learners.

**Figure 1:** A reading record for providing feedback on progress (date, book/page and remarks).

The iPod apps SuperWhy and MarthaSpeaks have been studied by the Rockman et al Company (2010), showing that they facilitate effective vocabulary acquisition. Nintendo's 'Brain Training' series has also been shown to be successful in a primary school classroom (Miller & Robertson, 2010). This suggests that the use of game-like tools in a primary school setting can be engaging and have educational benefits for pupils. Several commercial e-learning tools are also being developed to aid with children's reading, including BugClub and ReadingEggs. However, none of these examples currently include a robust mechanism for assessment in a way that is useful for a teacher. Other solutions, such as MindPlay, do include useful feedback systems but are typically self-contained. The implication being that it is not flexible enough to integrate into the existing culture at many primary schools.

## 2. Proposed Solution: The Vocalnayno Game

This paper proposes that a game could be developed to integrate into the existing culture and provide a new source of assessment for the teacher. Periodically, a parent or teaching assistant could provide a pupil an opportunity to play, which could be conducted on a laptop, tablet or similar device. While playing, the software measures the ability of the pupil to read phonics, letters, numbers and words aloud; saving the results to a database that is accessible to parents and teachers in addition to providing advice to helpers, as shown below in Figure 2:

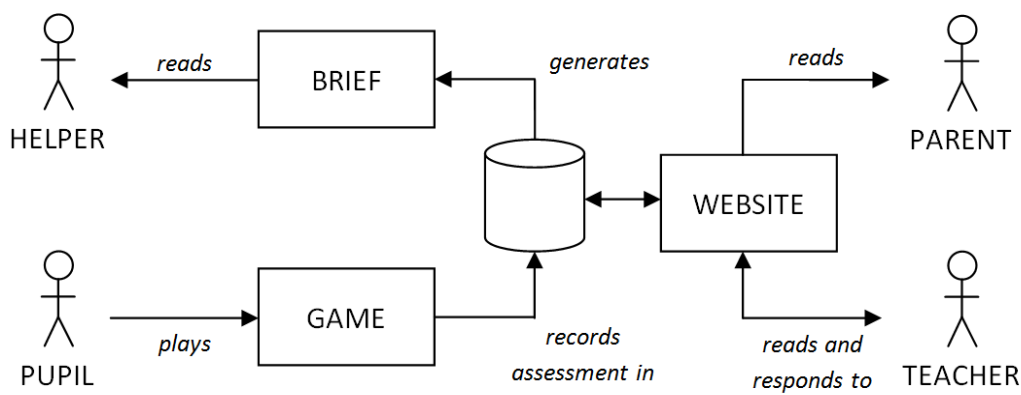

**Figure 2:** The intended approach for using Vocalnayno.

The player is presented with a story about an island being flooded. Inhabiting this island is the Vocalnayno character, a foam volcano who sits in the middle of the screen firing bubbles, as well as a number of 'primitive natives' who roam the surface. The goal of the game is to use the bubbles to capture as many of the native people as possible before the island is flooded. The basic gameplay is illustrated in Figure 3 below:

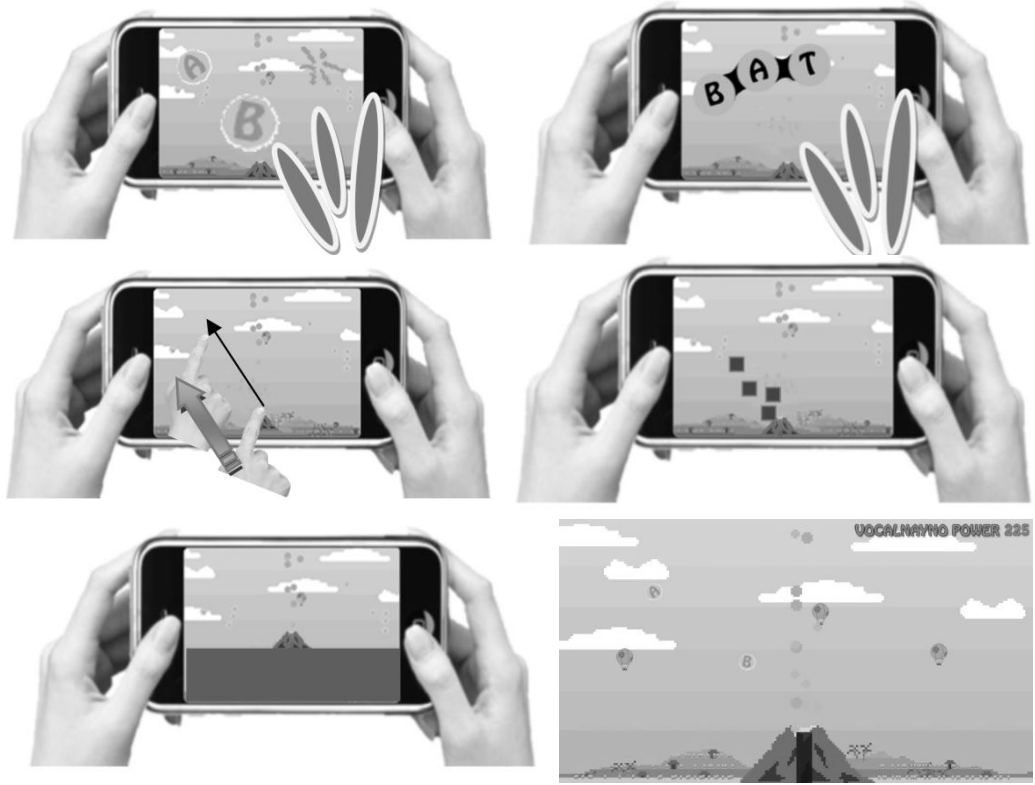

**Figure 3:** A brief storyboard of the Vocalnayno game

The game has two phases: powering up the Vocalnayno, followed by firing bubbles at the native characters. Power is a limited resource, accumulated by completing vocal exercises, which are periodically released as bubbles containing letters and words. Reading these aloud correctly increases the amount of power available to launch bubbles. The player then indicates the direction and speed they wish the bubbles to fire. When a bubble collides with a native character, they are sent into the sky, contributing to the score.

The main benefits of this game-based approach is that it blends a quantitative and objective mode of assessment, tailoring to the ability level of an individual pupil, with a potentially engaging and fun experience. It can also be made convenient for parents and teachers to access data about individual pupils by recording the results in a website-accessible database. Immediate concerns can be automatically flagged and prioritised, enabling teachers to provide advice to helpers more readily than the present paper-based approach. Assistants can also be prompted directly to respond to change as pupils become ready to progress to the next level of teaching material in other class activities.

## 3. Challenges, Lessons Learned & Future Work

In order to proceed with the project, a prototype was created using Multimedia Fusion Developer 2 (MMF2). Unfortunately, the initial results of a pilot test consisting of the author and one primary school pupil suggested that the native voice recognition capability of the chosen platform was not sufficient for the game to operate as intended. The author believes, however, that using a tailored voice recognition system based on a classification approach (e.g. Wu & Chan, 1993), testing for the specific on-screen phonetics, could provide a significant accuracy gain.

At the present time, no test deployment in a classroom environment has been conducted. However, this has led to an emphasis on requirements and design. Initial designs were presented informally to teaching assistants and parents, alongside children, at the end of the school day. Such informal discussions were important because they illustrated some of the challenges of garnering acceptance for a new technology in the classroom (Zhao et al, 2002), which had already manifested as concerns regarding the potential for misuse and detrimental impact. A design cannot be effective if the intended end-users are denied access to it. Since schools and teachers are accountable to parents and governors, the wider values of the design were situated within the context of the school's culture and operation. Consequently, this led the tool to evolve beyond a phonics acquisition supplement to supporting existing practice through providing assessment.

This form of consultation has also led to discussion about several other metrics that may be useful to teachers and teaching assistants. For example, measuring the challenges experienced and confidence gained by individual pupils. This could be achieved using eye-tracking in tandem with a front-facing camera, using a solution such as "EyePhone" (Miluzzo, Wang & Campbell, 2010) or similar (Yu et al, 2003; Goth, 2010), in order to measure the duration that phonic bubbles are gazed at. Furthermore, the relative volume at which a pupil reads aloud, as captured by a microphone, could be interpreted as an indication of confidence. Such data could supplement a record of individual ability, informing decisions about how to scaffold learning materials appropriately for each pupil.

Participants seemed happy to describe current practices at the school and discuss how they could be improved. It was convenient for each party to participate in this manner because of other commitments such as administration, afterschool clubs, work and childcare. Nevertheless, the disadvantage of this approach has been that not every stakeholder was involved in these discussions simultaneously, due to their informal nature.

## 4. Acknowledgements

Although redesigned for use in the educational context, Vocalnayno was originally conceived in collaboration with Allan Lowther, Douglas Pennant and Tomasz Kaczmarek at the Global Games Jam hosted by the International Game Developers Association in 2011. The author would like to thank them for their contribution while also acknowledging Justin Parsler, Douglas Brown and Professor Steve Jackson, who provided helpful feedback on an earlier version. In closing, a heartfelt thank you is extended to all of the anonymous stakeholders involved in the project.